\documentclass[ 
preprint,
superscriptaddress,
aip,
jcp,
 amsmath, amssymb,
floatfix,
]{revtex4-1}
\usepackage{bm}
\usepackage{graphicx}
\usepackage{textcomp}
\usepackage{array} 
\usepackage{float}
\usepackage{amsmath, amsthm}
\usepackage{mathrsfs}
\DeclareMathAlphabet{\mathpzc}{OT1}{pzc}{m}{it}
\usepackage{natbib}

\begin{document}
\renewcommand{\thefootnote}{\fnsymbol {footnote}}
\title{Neuroreceptor Activation by Vibration-Assisted Tunneling}

\author{Ross~D.~Hoehn}
\affiliation{Department of Chemistry, Purdue University,
West Lafayette, IN 47907 USA}
\author{David~Nichols}
\affiliation{Department of Medicinal Chemistry and Molecular Pharmacology, Purdue University,
West Lafayette, IN 47907 USA}
\author{Hartmut~Neven}
\affiliation{Google, Venice, CA 90291  USA }
\author{Sabre~Kais\footnote{Correspondance to: kais@purdue.edu}}
\affiliation{Departments of Chemistry and Physics, Purdue University,
West Lafayette, IN 47907 USA}
\affiliation{Qatar Environment and Energy Research Institute, Qatar Foundation, Doha, Qatar}

\begin{abstract}
G protein-coupled receptors (GPCRs) constitute a large family of receptor proteins that sense molecular signals on the exterior of a cell and activate signal transduction pathways within the cell. Modeling how an agonist activates such a receptor is fundamental for an understanding of a wide variety of physiological processes and it is of tremendous value for pharmacology and drug design. Inelastic electron tunneling spectroscopy (IETS) has been proposed as a model for the mechanism by which olfactory GPCRs are activated by a bound agonist. We apply this hyothesis to GPCRs within the mammalian nervous system using quantum chemical modeling. We found that non-endogenous agonists of the serotonin receptor share a particular IET spectral aspect both amongst each other and with the serotonin molecule: a peak whose intensity scales with the known agonist potencies. We propose an experiential validation of this model by utilizing lysergic acid dimethylamide (DAM-57), an ergot derivative, and its deuterated isotopologues; we also provide theoretical predictions for comparison to experiment. If validated our theory may provide new avenues for guided drug design and elevate methods of \emph{in silico} potency/activity prediction.
\\*
\\*
{\center
\large{\textbf{Significance Statement}} \normalsize}
\\*
This project provides a scheme by which the potency of small molecule drugs at their target transmembrane receptor may be predicted by a recent theoretical model. This is a vital task as the fundamental mechanism by which receptor proteins are activated via quantum mechanical processes.
\end{abstract}

\date{\today}

\maketitle

\section*{Introduction}
Quantum activity within biological systems and the applications of information theory therein have drawn much recent attention \cite{Arndt, Davies, snp1, Plenio2, Rab1}. Examples of systems that exploit such phenomenon are: quantum coherence and entanglement in photosynthetic complexes \cite{newforSK, 1367-2630-13-11-115002, snp2, snp5, snp6, snp7, AAG2, Rab2, Schol1, Schol2}, quantum mutations \cite{McFadden, Khalili}, information theory and thermodynamics of cancers \cite{Levine1, Levine2},the avian magnetic compass \cite{10.1371/journal.pone.0000937, snp3, snp4, Plenio1}, tunneling behavior in the antioxidant breakdown of catechols present in green tea \cite{doi:10.1021/ja063766t}, enzymatic action\cite{Garcia}, olfaction\cite{TURIN2002367}, and genetic coding \cite{Patel}. G Protein-Coupled Receptors (GPCR) are the target for the greatest portion of modern therapeutic small molecule medications\cite{Christopoulos}. Predictability of pharmacological efficacy and potency for new drugs prior to a complex total synthesis can be aided by \emph{in silico} methods such as pharmacophore modeling, or the construction of homology models. Protein/agonist binding theory has been described through variants of the “Lock and Key” model, originally proposed by Fischer\cite{Fischer}, and the extensions thereof\cite{Koshland}. Although this theory has provided insights into free energy changes associated with the formation of the activated protein/agonist complex, it has not manifested sufficient capacity for the prediction ligand potency due to a lack of knowledge concerning the mechanism by which the agonist activates the complex.  Growth of the computational power of modern machines as well as developments within the field of computational chemistry and molecular modeling has afforded reconnaissance and scouting methods in the field of drug design.  Additionally methods such as QM/MM have been used in studies of protein folding and generation of the protein-agonist activated complex.  Furthering our ability to predict information regarding the viability of molecules as drugs is greatly important.

Early models attempting to account for predictability of agonist classification beyond mere shape were those of odorant binding\cite{Dyson, Wright1977473}; these works proposed a vibrational theory of receptor activation. Vibrational theories were eventually disregarded for reasons that include a lack of conceived mechanism and the inability of the protein (which undertakes thermally driven random walks in their conformation) to detect the continuum of thermally-activated, classical vibrations of the odorant. A recently suggested theory for olfactory activation consists of a physical mechanism closely resembling Inelastic Electron Tunneling Spectroscopy (IETS) \cite{TURIN2002367, Turin01121996, Gustation}. The plausibility of time scales associated with this process was verified to be consistent with relevant biological time-scales through Marcus theory\cite{PhysRevLett.98.038101}. Electron tunneling rates for the olfaction system have been calculated and support the theory \cite{NP8}. Furthermore, eigenvalue spectral analysis of odorant molecules has shown a high correlation between the vibrations and odorant classification \cite{NP7}.

We focus on an initial examination of the viability of the vibrational theory of protein activation in cases involving protein-agonist binding within the central nervous system via application of IETS theory as a predictor of potency as defined within \cite{Urban01012007}. Activation of the 5-HT$_{1A}$ and 5-HT$_{2A}$ receptors is implicated as being associated with human hallucinogenic responses \cite{Nichols2004131, Serotonin, Morena}. We utilize a model of inelastic electron tunneling to describe the protein/agonist complex in a manner that will utilize the vibrational frequencies of the bound agonist to facilitate electron transfer within the activation site of the protein/agonist complex. The prerequisite agonist information was collected through molecular quantum mechanics calculations utilizing density function theory as well as normal mode analysis and natural bonding order methods; necessary were the harmonic displacements, frequencies and partial charges of each constituent atom. In Section II, we will first present a qualitative discussion of the relationship between the tunneling model and the protein-agonist complex. Section III will discuss the tunneling features of several 5-HT$_{1A}$ and 5-HT$_{2A}$ agonists, and how these correlate with the potency of these molecules taken from previous studies \cite{Parrish, Barden}. We conclude with a proposed set of molecules that could be employed in experimental validation of the vibrational theory's applicability in the central nervous system and present the expected results in accordance with this proposed mechanism.

\section*{Mapping the Model into the Biological System}

Application of the IETS model for the agonist’s protein environment requires mapping several aspects of the IETS methodology into the biological system. The two-plate apparatus of the tunneling junction herein represent the walls of the receptor site; more explicitly, under electron transfer, the valance and conductance bands associated with each side of the junction are mapped to specific HOMOs and LUMOs of particular residues comprising the walls of the receptor. This dictates that energy transition detectable by the protein should be the energy difference between electronic levels of residue side-chains or any bound cofactors, such as a metal ion. This alteration of IETS also localizes the source of tunneling electrons to a single residue side-chain; the implication is that electrons are not capable of distributed tunneling through the molecule in the manner of a doped analyte within a junction. This lack of a spatial distribution of electron trajectories suggests that the act of tunneling is localized to regions of the agonist molecule along the classical trajectory of the tunneling electron between the site of the electron donor and the site of the electron acceptor. This implies that not all local oscillators associated with a specific mode may fully contribute to the current enhancement due to the fall-off of the charge-dipole  coupling between the tunneling electron and the local atomic oscillators.

Secondly, unlike the typical experimental IETS procedure, the analyte is not deposited upon a surface, being encapsulated by the active site. There is no externally applied potential within the receptor site which would have allowed for the scanning of energies; yet, it has been suggested that an ionic cofactor, likely a calcium ion, could provide this driving field. The implication of this is that the receptor is set to test the vibrational-assisted enhancement to the electron tunneling rate at a specific energy, opposed to scanning a range of energies. The electrostatic interactions which govern docking orientation would be a means of orienting the endogenous agonists in such a way that the tunneling junction is appropriately aligned for maximized electron transfer across the atoms responsible for the inelastic contribution. Non-endogenous agonists would align with residues in a manner which may place energetically appropriate vibrational modes of the agonist in proximity of the tunneling vector specific to this protein, thus allowing for the activation of the receptor.

\section*{Results}
\label{Sec:D-Scaling}
Generation of tunneling spectra was completed through the procedure described by Turin \cite{TURIN2002367, 0022-3719-19-33-013}, and outlined within the Supplementary Material. This procedure is an adaptation of earlier inelastic tunneling literature\cite{doi:10.1080/13642818508240633, PhysRevB.14.3177} and similarly uses arbitrary units (\emph{a.u.}) for the tunneling intensity. Our spectral procedure was validated by comparison of the spectra of the formate ion, which is prevalent throughout experimental and theoretical literature in IETS \cite{0022-3719-19-33-013}. These \emph{a.u.} are proportional to the conductance enhancement, as well as representative of an enhancement to the magnitude of electrostatic charge-dipole interaction an electron experiences during tunneling. Necessary information for implementing the calculations - outlined in the Supplementary Material - was collected through quantum chemical calculations. Computations were performed using Density Functional Theory at the 6-311G basis-level, utilizing the B3LYP functional which serves well for organic hydrocarbons; in contrast to similar previous works\cite{TURIN2002367, 0022-3719-19-33-013}. Expanded pseudopotential correlation consistent 5-zeta basis was used for large atoms where necessary \cite{Dunning}. DFT was employed both due to its high accuracy in transition dipole frequencies and due to a desire to avoid encroaching errors associated with dissimilarities between analyte and parameter molecules in semi-empirical methods. Initial applications of Hartree-Fock theory displayed the characteristic 0.8 factor shift to the vibrational frequencies, which is less than desirable for ease of interpretation of the vibrational spectra. Vibrational calculations utilize reduced modal displacements, $\mu$; proportional to the Cartesian displacement through the modal center-of-mass factor, $\sqrt{\mu}$. This factor arises due to use of center-of-mass coordinates within the classical theory after application of the harmonic approximation during calculations of the normal modes. Natural bond order calculations were performed to yield the partial charges, $q_{i}$ attributed to each atom constituting the agonist. Scaled Kronecker delta functions are plotted at the on-resonance absorbance frequency of the mode; these functions were convolved with Gaussian functions possessing a conservative FWHM of 25 cm$^{-1}$, representing a very narrow thermal distribution. The spectral width was introduced to allow for peak additions, while 25cm$^{-1}$ was selected to avoid over estimations of peak breadth.

Assessment of vibrational bands from the 5-HT$_{2A}$ agonists which could facilitate the inelastic transfer of electrons within the protein environment is of primary import. Agonists of a particular protein would share a single spectral feature associated with the inelastic transfer, as the same amino acid side-chains would be responsible for the electron donation and acceptance for a specific protein. Tunneling spectrum of several selected 5-HT$_{2A}$ agonists have been generated. LSD, was selected as it possesses a high potency as well as activity at this particular serotonin receptor within the cortical interneurons\cite{NP9}. DOI (2,5-dimethoxy-4-Iodo-amphetamine) was selected due to its high selective for the 2A-subtype receptor \cite{NP10}. The remaining selected molecules are members of the 2C-X (4-X-2,5-dimethoxyphenethylamine) class of psychedelic phenethylamines. All compounds selected are known hallucinogens \cite{vbn4a, vbn4b, vbn4c} some first characterized by Alexander Shulgin in the compendia works PiHKAL and TiHKAL\cite{PiHKAL, TiHKAL}.

\begin{figure}
\centering
\includegraphics[scale=.70]{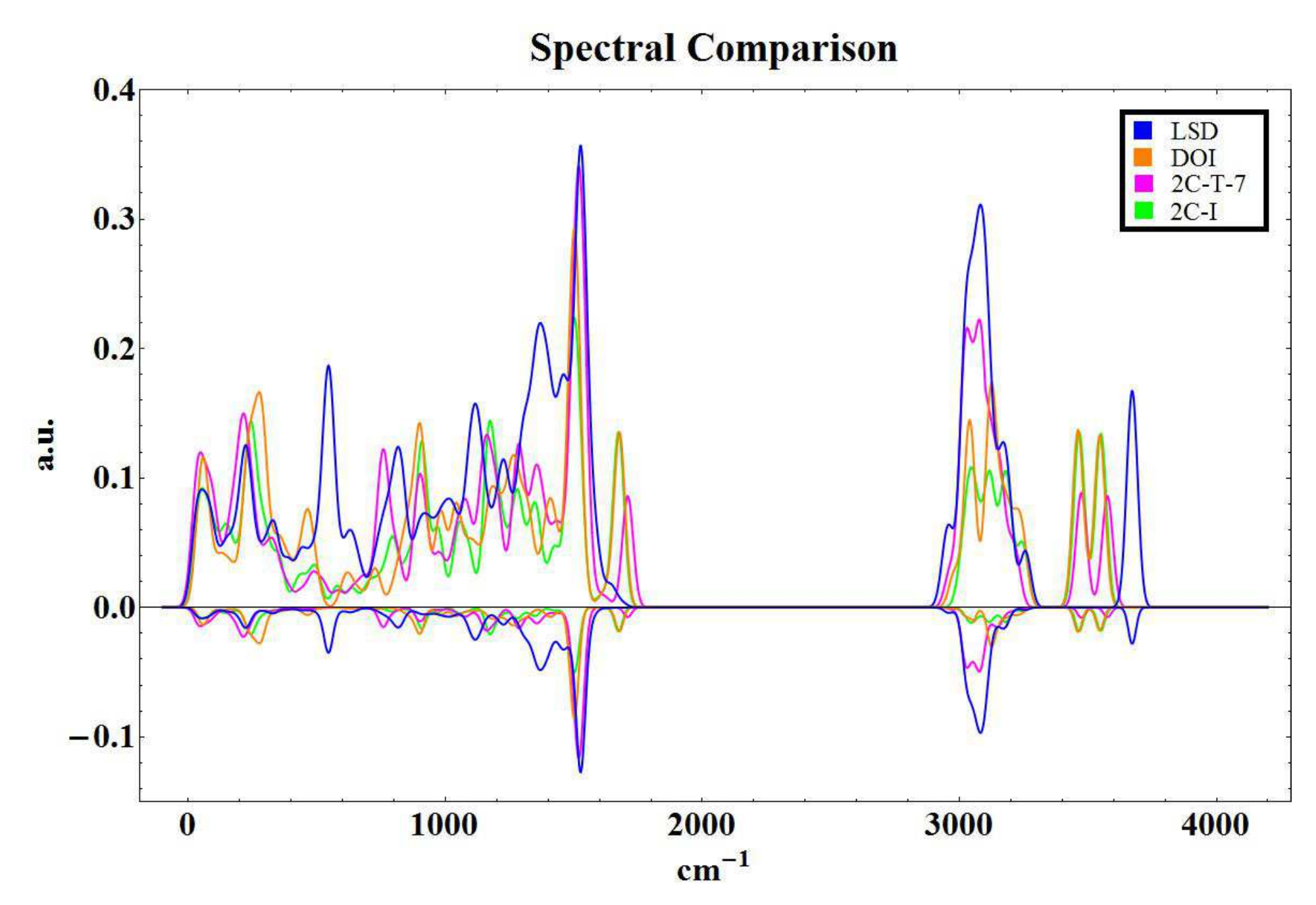}
\caption{The tunneling spectrum of several known 5-HT$_{2A}$ agonists as well as the square of the tunneling PDF reflected below the energy axes; the square is used to highlight major spectral aspects. The Spectral Similarity Index of each plot given in the inlay, noting that these similarity indices allude to good spectral agreement with the reference spectrum, LSD.  More detailed information is provided within the Supplemental Material}
\label{fig:Spectral_Simularity}
\end{figure}

Figure \ref{fig:Spectral_Simularity} shows the tunneling spectra of select agonists (above the axis). The selection of candidate peaks, possibly responsible for facilitating inelastic transfer, was performed using the Spectral Similarity Index (SI), similar to that used for comparison of mass spectra \cite{NP12}. The SI was taken over the entire spectral region and repeated for a scan of the local regions with an overlapping step of 500cm$^{-1}$ with a width of 1000cm$^{-1}$. The SI is given by:

\begin{equation}
SI = 1 - \sqrt{\frac{ | a_{i} - b_{i} | }{N}}
\end{equation}

Where, within the above equation, $N$ is the normalization constant (the numerator performed for spectra $b$ and $a=b$); $b_{i}$ is the value of the spectra being analyzed at discrete location $i$ while $a$ is a reference spectra. Being the most potent agonist, LSD was selected as the reference spectra for our SI calculations. The SI’s, both global and local scans, associated with each of the tunneling spectra can be found within a table provided in the Supplementary Material. To highlight major aspects of the tunneling PDF, we squared the function, exaggerating aspects which exhibit large tunneling amplitudes within the spectra (Figure \ref{fig:Spectral_Simularity} reflected below energy axis). The only broadly shared spectral aspects were those at ~1500cm$^{-1}$. For a more thorough discussion of the spectral aspects, isotopic effects at functional groups and density of states for these systems, please see the provided Supplementary Materials.

The integral of the tunneling probability density was taken around the 1500 $\pm$ 35cm$^{-1}$ region and compared to known EC50 data for compounds shown to be agonists of the 5-HT$_{2A}$ receptor.  The EC50 used within this paper is taken from Parrish et al \cite{Parrish} who determined the elevated levels of phosphoinositides associated with the activation of the 5-HT$_{2A}$ receptors of the human A20 cell line employed within the experiement across a collection of compounds from the phenethylamine (PEA, or 2C-X) and phenylisopropylamine (PIA, or DOX) classes.   The detection procedure was replicated from Kurrasch-Orbaugh et al. \cite{ Kurrasch}.  The selection of data used for our comparison is provided within our Supplemental Material.  Data was selected from a single source, helping to assure uniformity in collection and determination, while selecting and comparing members of specific families of molecules (i.e. PIA/PEAs) helps to minimize drastic changes in their docking configurations which may affect potency.  The similarities granted by selecting compounds from families may not allow for substantive prediction in the relative potency, beyond docking affinities; this can be seen by noting that two PIAs, DOI and DOB, have similar docking affinities at the 5-HT$_{2A}$ receptor \cite{Ray}, while possessing great differences in their potency \cite{Moya}.

\begin{widetext}
\begin{figure*}
\centering
\includegraphics[scale=.25]{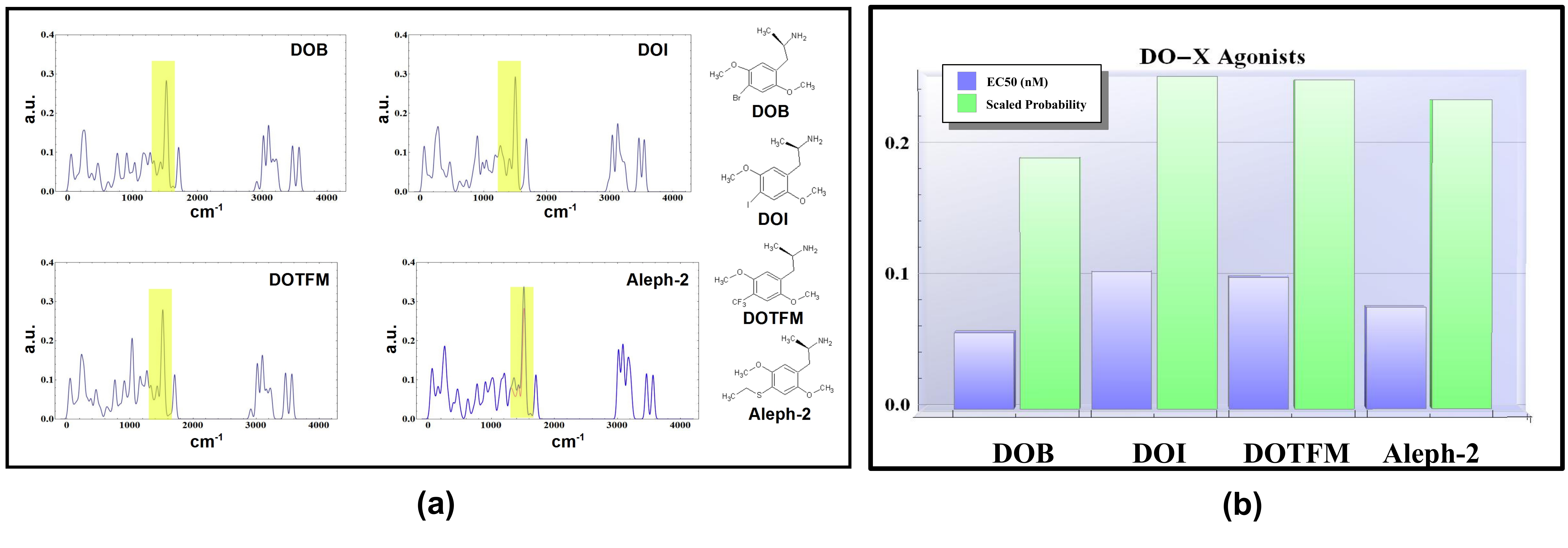}
\caption{(a) The tunneling spectra of several DOX class agonists as well as their molecular structures. (b) The inverse of the median effective concentration for the DOX class agonists plotted against the tunneling probability within the region at 1500$\pm$35cm$^{-1}$. The trend of tunneling intensity follows roughly the trend of the agonist's potency at the 5-HT$_{2A}$ receptor.}
\label{fig:FullDOFigure}
\end{figure*}
\end{widetext}

\begin{widetext}
\begin{figure*}
\centering
\includegraphics[scale=.25]{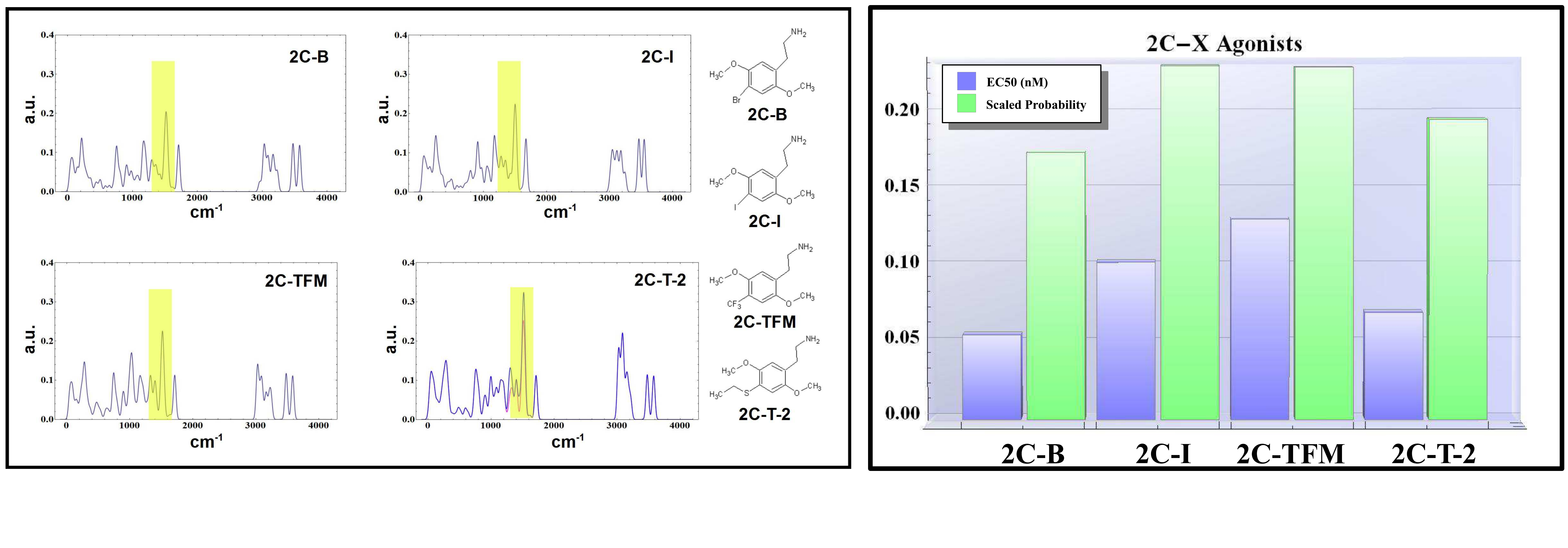}
\caption{(a) The tunneling spectra of several 2C-X class agonists as well as their molecular structures. (b) The inverse of the median effective concentration for the 2C-X class agonists plotted against the tunneling probability within the region at 1500$\pm$35cm$^{-1}$. The trend of tunneling intensity follows roughly the trend of the agonist's potency at the 5-HT$_{2A}$ receptor.}
\label{fig:Full2CFigure}
\end{figure*}
\end{widetext}

The effective concentrations of several phenethylamines were taken from \cite{Parrish} and compared to the local integrals of the tunneling PDF. This comparison exposes a possible correlation to the inverse of the EC50 data, taken to be representative of the potency for each species at the receptor. Results for the 1500cm$^{-1}$ region are shown in Figures \ref{fig:FullDOFigure} and \ref{fig:Full2CFigure} for the DOX class and 2C-X class molecules computed, respectively. Figures \ref{fig:FullDOFigure}a and \ref{fig:Full2CFigure}a give the tunneling spectra for each molecule, Fig. \ref{fig:FullDOFigure}b and \ref{fig:Full2CFigure}b compare the integral values to the known EC50s.

As inelastic tunneling facilitated by a charge-dipole interaction is a highly local process where the interaction potential falls-off as r$^{-3}$ for non-parallel displacements. Modes not local to the electron donor/acceptor sites will not maximally contribute to the electron transfer, which is proposed to be responsible for protein activation. Particular modes in 2C-T-2 and in Aleph-2 reside within the thioether (roughly 5 angstrom from the ring system); due to the non-locality of these oscillators, tunneling probability should be examined after having removed their contributions from the spectra. Figures \ref{fig:FullDOFigure}a and \ref{fig:Full2CFigure}a present the tunneling spectrum of 2C-T-2 and Aleph-2 disregarding these contributions. After correction for non-local motions, the integrals are in good qualitative agreement with the inverse EC50.This preliminary information supports a possible quantum mechanical origin for the activation of sensory proteins. We shall propose a possible experimental validation of the theory within the following section. 

\section*{Proposed Experiment}
\label{Experi}

Early findings suggest that both the lake whitefish and the American cockroach can identify isotopologues of amino acids and pheromones, respectively \cite{np4, np5}. Recent experiments have shown that the common fruit fly presents both naive bias to and a potential for trained aversion towards the isotopologues of acetophenone \cite{Franco14022011}, and reposte \cite{NP3}. Recent works featuring human subjects have shown that naive participants are incapable of discerning between deuterated acetophenone\cite{NP6}; a second study\cite{NP2}presented evidence which suggests human capability at discerning deuterated variants of musk odorants. Other works have attempted to identify the characteristic vibrations associated with particular odors\cite{NP1}, yet have not explicitly considered an electron tunneling mechanism.  In this spirit, we propose an experimental procedure for testing this new iteration of the vibrational theory of protein activation \emph{in vivo}.

DAM-57 (N,N-dimethyllysergamide) is an ergot derivative with a mild hallucinogen effect associated with activity at the 5-HT$_{2A}$ receptor. As it activates the same receptor, the above discussed candidate peak should, and does, appear in the tunneling spectrum of DAM-57.  By using isotopologues of a single compound, we may minimize any differences which may lead to alterations in either docking geometry or affinity for the activation site.  It, however, should be noted that binding isotope effects and kinetic isotope effects can cause differences in the potency of compounds, but this effect rarely outstrips 10\% \cite{Swiderek} so any effect on the order of 30

\begin{widetext}
\begin{figure*}
\centering
\includegraphics[scale=.20]{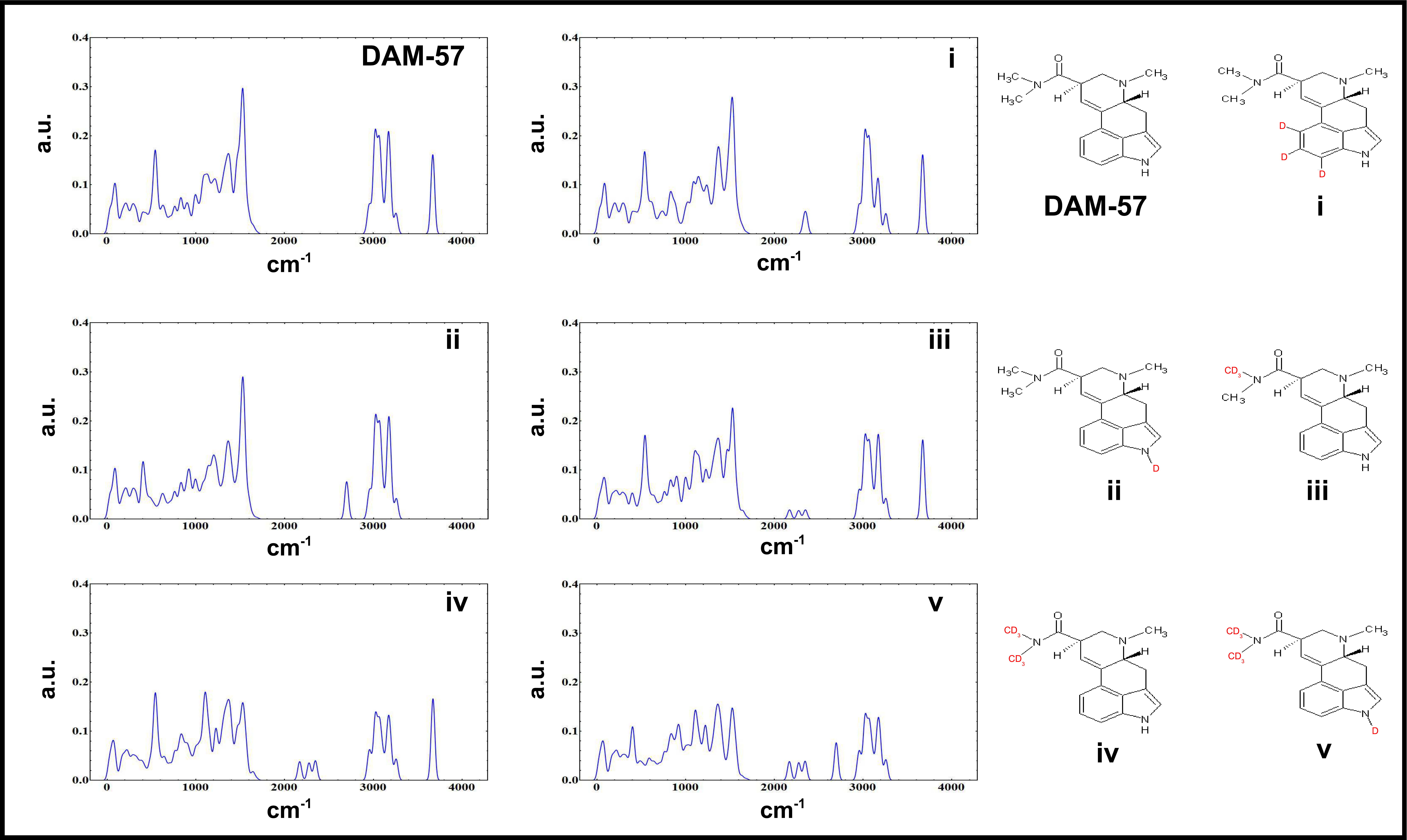}
\caption{The tunneling spectrum of several deuterium-isotopologues of DAM-57. Yellow highlights have been given to the energy region which is assumed to be the active energy region for inelastic tunneling transfer. Specific dueterations deplete the tunneling probability within this region, and may effectively eliminate the agonism of the molecule within the 5-HT$_{2A}$ receptor.}
\label{fig:DAMClassPlots}
\end{figure*}
\end{widetext}

Using 1500 cm$^{-1}$ as a central point, and recalling the applied FWHM, the modes contributing to inelastic transfer are those at 1500 $\pm$35 cm$^{-1}$. Modes within that range have motions associated with (in order of contribution): stretching of the amide methyl hydrogen; stretching of the phenyl and indole hydrogens; and bending of the methyl hydrogen of the tertiary amine. 

Deuteration of the three phenyl hydrogens (DAM-57-i) yields a marginal attenuation in intensity near 1500 cm$^{-1}$, and small change in tunneling probability. DAM-57-ii displays a reduction in the ~3700 cm$^{-1}$ region, N-H stretch, shifting weight to ~2700 cm$^{-1}$. Deuteration of the indole amine results in almost no character change near the active region. Pro-deuteration of a single amide methyl (DAM-57-iii) significantly decreases the tunneling intensity in the 1500 cm$^{-1}$ region. Continued deuteration of the amide system (DAM-57-vi), reduces this peak to roughly one-half the pro-protium intensity. DAM-57-vi and DAM-57-v, moiety co-deuteration scenarios, present very small alterations of the peak intensity when compared to DAM-57-iii and DAM-57-vi, respectively.

Within the tunneling model, deuteration of the amide side chains should dampen the activity of the molecule at the 5-HT$_{2A}$ receptor through a relative reduction in potency. This conclusion is supported by the relative activity between DAM-57 and LSD. The flexible ethyl amide of LSD has been found to be essential to its high activity \cite{Serotonin, B906391A, Nichols2004131, BayarA2005225}, and that the methyl analogue (DAM-57) is far less potent; the tunneling probability at the ~1500 $cm^{-1}$ region of DAM-57 is depleted when compared to that of LSD. Following this, a prediction that further depletion of the tunneling probability within this region should continue to diminish the potency at the receptor may be entertained. The intensity of the tunneling spectrum of DAM-57-iv is roughly a third the pro-protium, and the probability density of tunneling is roughly tenthed – this implies a possible extreme loss of potency associated with deuteration of the amide side-chains.

\section*{Conclusions}
Herein we describe the agonist-protein system by an electron tunneling junction coupled to a field of oscillating dipoles, representative of the constituent atoms of the agonist. The oscillator field provides a secondary path for electron transfer between the donor and acceptor states of the junction. This secondary inelastic path facilitates the transfer if and only if the electron can donate a quantum of energy to the oscillator field. Using this method we examined classes of agonists for the 5-HT$_{2A}$ receptor and found that all agonist, to varying degrees, are capable of facilitating electron transfer within the same energy region. The degree to which this tunneling is facilitated correlates roughly to the potency of the agonist within our test cases. We examined the tunneling characteristics of isotopologues of these agonists and predict that it may be possible to modulate or quench their agonist properties though the isotope exchange of specific atoms. Also included is a proposed experimental path to test the model described herein. We conclude that this mechanism is a candidate for the activation step for some transmembrane proteins, and its examination may allow for better prediction of candidate drug molecules and the possible ability to control agonism of molecules.

\section*{acknowledgements}

This work is supported by the NSF Centers for Chemical Innovation: Quantum Information  for Quantum Chemistry, CHE-1037992

\bibliography{Bibio}
\end{document}